\documentclass[modern]{aastex631}

\newcommand{\sdo}{{\it SDO}} %
\newcommand{\ic}{$I_\mathrm{c}$}%
\newcommand{\ilc}{$I_\mathrm{lc}$}%
\newcommand{\veight}{$V_\mathrm{80}$} %
\newcommand{\vsix}{$V_\mathrm{60}$} %
\newcommand{\vfour}{$V_\mathrm{40}$} %
\newcommand{\vtwo}{$V_\mathrm{20}$} %
\newcommand{\vhmi}{$V^\mathrm{HMI}_\mathrm{like}$} %
\newcommand{\vcog}{$V_\mathrm{CofG}$} %
\newcommand{\dphi}{$\delta\phi$} %
\newcommand{\dphiI}{$\delta\phi(I_\mathrm{lc} - I_\mathrm{c})$}
\newcommand{\dphiVL}{$\delta\phi(V - I_\mathrm{lc})$}
\newcommand{\dphiVI}{$\delta\phi(V - I_\mathrm{c})$}
\newcommand{\dt}{$\delta\tau$} %
\newcommand{\FeI}{\ion{Fe}{1}\,6173.3\,\AA} %
\newcommand{\kmps}{km\,s$^{-1}$} %
\newcommand{\mps}{m\,s$^{-1}$} %
\newcommand{\nuac}{$\nu_\mathrm{ac}$} %
\newcommand{\tab}{$\tau_\mathrm{AB}$} %
\newcommand{\dtab}{$\delta\tau_\mathrm{AB}$} %
\newcommand{\tba}{$\tau_\mathrm{BA}$} %
\newcommand{\tabp}{$\tau_\mathrm{AB^{\prime}}$} %
\newcommand{\tbbp}{$\tau_\mathrm{BB^{\prime}}$} %
\newcommand{\tbpa}{$\tau_\mathrm{B^{\prime}A}$} %

\shorttitle{Phase Shifts in Helioseismic Waves}
\shortauthors{Zhao et al.}

\begin{document}
\title{Phase Shifts Measured in Evanescent Acoustic Waves above the Solar Photosphere and Their Possible Impacts to Local Helioseismology}
\author[0000-0002-6308-872X]{Junwei Zhao}
\affiliation{W. W. Hansen Experimental Physics Laboratory, Stanford University, Stanford, CA 94305-4085, USA}

\author[0000-0003-0003-4561]{S.~P.~Rajaguru}
\affiliation{Indian Institute of Astrophysics, Bangalore-34, India}

\author[0000-0002-2632-130X]{Ruizhu Chen}
\affiliation{W. W. Hansen Experimental Physics Laboratory, Stanford University, Stanford, CA 94305-4085, USA}

\begin{abstract}
A set of 464-min high-resolution high-cadence observations were acquired for a region near the Sun's disk center using the Interferometric BI-dimensional Spectrometer (IBIS) installed at the Dunn Solar Telescope. 
Ten sets of Dopplergrams are derived from the bisector of the spectral line corresponding approximately to different atmospheric heights, and two sets of Dopplergrams are derived using MDI-like algorithm and center-of-gravity method. 
These data are then filtered to keep only acoustic modes, and phase shifts are calculated between Doppler velocities of different atmospheric heights as a function of acoustic frequency. 
The analysis of the frequency- and height-dependent phase shifts shows that for evanescent acoustic waves, oscillations in the higher atmosphere lead those in the lower atmosphere by an order of 1\,s when their frequencies are below about 3.0\,mHz, and lags behind by about 1\,s when their frequencies are above 3.0\,mHz. 
Non-negligible phase shifts are also found in areas with systematic upward or downward flows.
All these frequency-dependent phase shifts cannot be explained by vertical flows or convective blueshifts, but are likely due to complicated hydrodynamics and radiative transfer in the non-adiabatic atmosphere in and above the photosphere. 
These phase shifts in the evanescent waves pose great challenges to the interpretation of some local helioseismic measurements that involve data acquired at different atmospheric heights or in regions with systematic vertical flows. 
More quantitative characterization of these phase shifts is needed so that they can either be removed during measuring processes or be accounted for in helioseismic inversions.
\end{abstract}


\section{Introduction}
\label{sec1}

Helioseismology investigates the Sun's interior structure and dynamics through analyzing the oscillation signals observed in either Doppler velocities or intensities in or above the Sun's photosphere \citep{jcd02}. 
Despite great advances and breakthroughs obtained in the last few decades using both global and local helioseismological analysis methods, helioseismology still faces challenges of some systematic effects, whose presence in either data or analysis methods often complicates the interpretation of measurements or hinders a reliable inference of the Sun's interior properties. 
For instance, the helioseismic center-to-limb effect \citep{zha12} is an order of magnitude larger than the meridional-circulation-induced signals for certain travel distances \citep{zha13, kho14, raj15, che18}, posing a great challenge to our inferences of the Sun's deep meridional circulation.
Another systematic effect occurs in and near magnetic regions. 
As demonstrated by \citet{giz09}, the inferences of subsurface sound-speed perturbations using different helioseismic analysis techniques gave sharply different results, implying a systematic effect in such regions. 
To reduce this effect, \citet{lia15} and \citet{che17} excluded all the magnetic regions with a strength above a certain threshold.

What causes the center-to-limb effect and the magnetic effect in the helioseismic analysis is not exactly known, but recently \citet{zha20} suggested that the center-to-limb effect and part of the magnetic effect might have a similar cause: the helioseismic signals used in the analysis, particularly those pairs of signals used to compute the cross-correlation functions in time--distance helioseismology, are not observed in the same atmospheric heights. 
This height difference may introduce unexpected phase shifts in acoustic waves.
Actually, the possibility that the helioseismic center-to-limb effect may be related to different heights where oscillatory signals are taken has already been explored by various authors. 
In the photosphere, convective blueshifts in granules dominate in areas over redshifts, systematically shifting the phases of acoustic waves and causing the observed center-to-limb effect according to \citet{bal12}; but the shift caused by this process is not expected to be frequency dependent and the values of shifts are expected be small. 
Through coupling the Doppler and intensity observations, \citet{sch15} demonstrated how granulation can change the observed amplitude and phases of the oscillatory signals. 
However, how the phases of the oscillatory signals vary with atmospheric height and acoustic frequency has not been well studied, although a detailed analysis of the behaviors of these waves above the photosphere is expected to offer a key insight of the physical causes of the observed effect.

Meanwhile, we have to recognize that most of the helioseismic waves that we observe are evanescent waves.
Although helioseismic waves are observed between approximately 2.0 and 8.0\,mHz, the photospheric cutoff frequency of the waves is believed near about 5.0\,mHz \citep{jim11}, meaning that waves with frequencies below 5.0\,mHz, including the strongest oscillatory power near 3.0\,mHz, are evanescent waves, and those with frequencies above 5.0\,mHz are propagating waves. 
While it is often expected that phases of evanescent waves no longer change with height in the atmosphere unless perturbed by flows or other physical factors, whether this is exactly the case in observed intensities and Doppler velocities remains to be examined. 
Despite the early effort examining how phases change with height \citep{lit79}, similar efforts with a higher precision that meets the current helioseismic measurement requirements have not been done, likely due to the scarcity of high-quality simultaneous observations covering multiple atmospheric heights. 

In this article, through analyzing a set of well-observed long-duration data with high spatial and spectral resolutions, we study how the phases of both evanescent and propagating waves change with atmospheric height. 
We believe that our results have a profound impact on the interpretation and inversion of many local and global helioseismic measurements, which may challenge us to design new methods removing these systematic phase shifts. 
This article is organized as follows: we introduce our data acquisition and reduction in Section~\ref{sec2}, present our measurements of various types of phase shifts in Section~\ref{sec3}. We then discuss our results in Section~\ref{sec4}, and give conclusions in Section~\ref{sec5}.

\section{Observation and Data Reduction}
\label{sec2}

\begin{figure}[!t]
\centering
\includegraphics[width=1.0\textwidth]{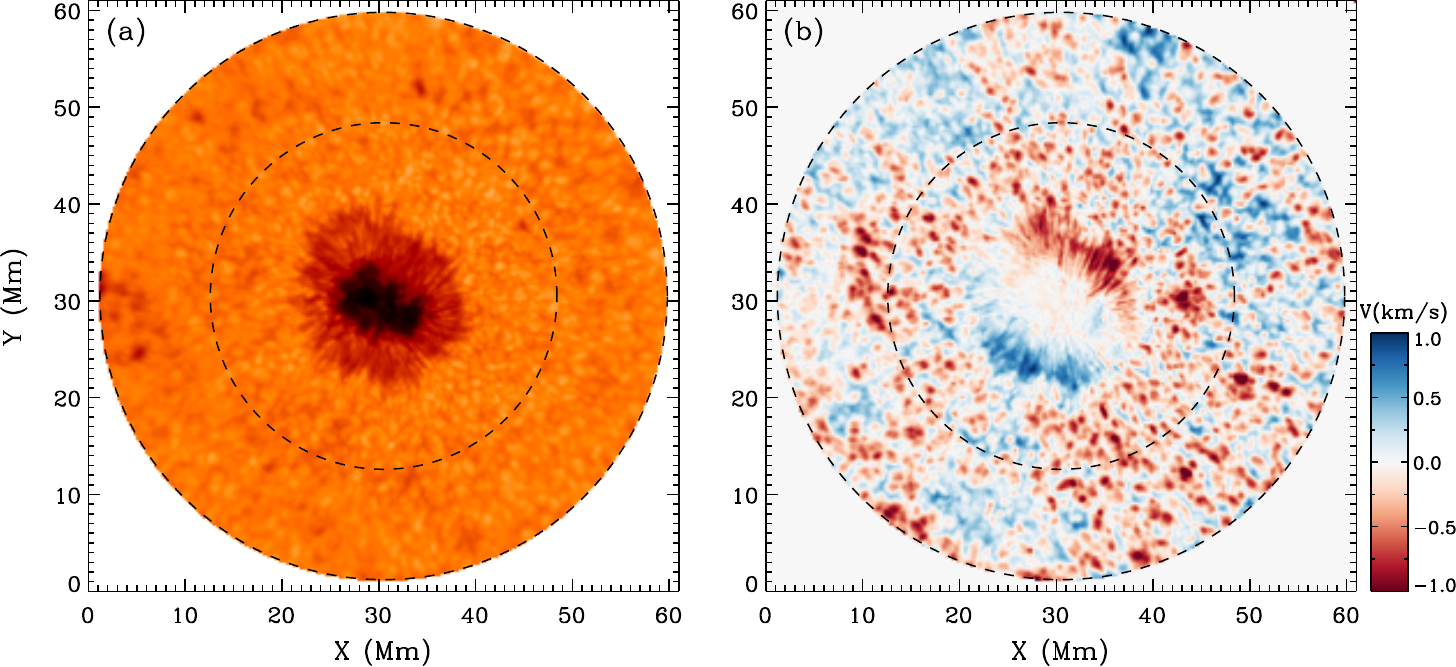}
\caption{(a) A continuum image showing the field of view of the observation with a sunspot at the center of the field. (b) A sample image of Doppler velocities \vfour, with blue representing blueshift and red representing redshift. The area between the two dashed circles in both panels is used as quiet-Sun region in this study.}
\label{img}
\end{figure}

The data used in this study were previously acquired and used for different studies \citep{raj10, cou12}. 
These imaging spectropolarimetry data were obtained on 2007 June 8 using the Interferometric BI-dimensional Spectrometer \citep[IBIS;][]{cav06} installed at the Dunn Solar Telescope at Sacramento Peak, New Mexico.
These observations have a spectral resolution of 25\,m\AA, with all the Stokes parameters ($I, Q, U, V$) taken at 23 line positions along the spectral line \FeI.
The spatial resolution of this set of observations is $0\farcs330$ (or $0\farcs165$ pixel$^{-1}$), and the circular field of view has a diameter of $80\arcsec$. 
The observed region is close to the disk center at S07W17, with a medium-size sunspot of NOAA AR 10960 near the center of the field (see Figure~\ref{img}a). 
The temporal cadence of the observation is 47.5\,s, and a total of 586 time steps were taken with a duration of approximately 464\,min. 
The \FeI\ line, which is also used by the Helioseismic and Magnetic Imager onboard {\it Solar Dynamics Observatory} \citep[\sdo/HMI;][]{sch12a, sch12b} and by the Polarimetric and Helioseismic Imager onboard {\it Solar Orbiter} \citep{sol20}, is believed to form between $\sim$100--270\,km above the photosphere according to various authors \citep{nor06, fle11, kit15}. 
The spectral images were then dark subtracted, flat-fielded, and re-registered to remove the atmospheric distortions that were derived from the white-light images recorded simultaneously \citep{raj10}. 

\begin{figure}[!t]
\centering
\includegraphics[width=0.55\textwidth]{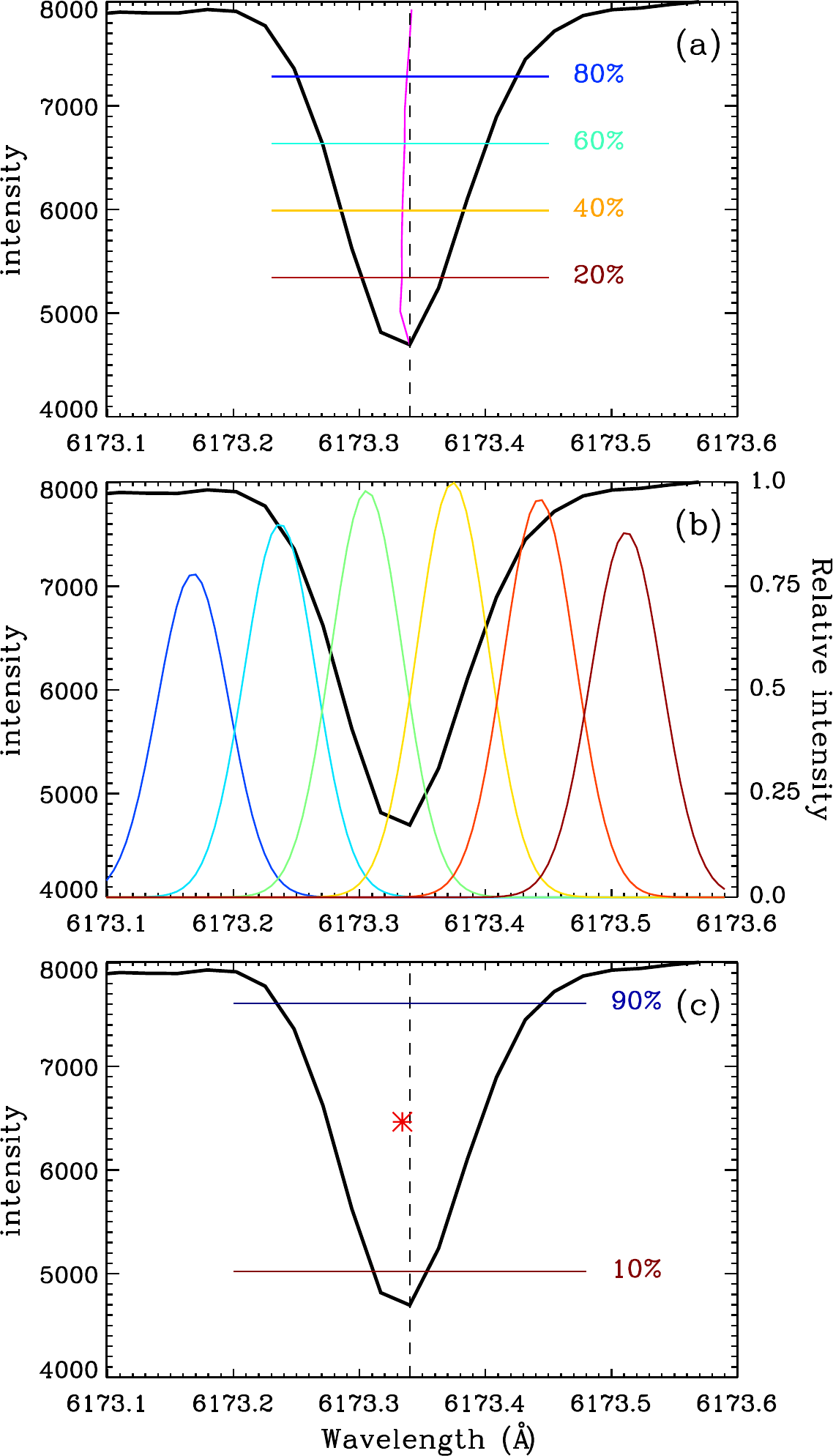}
\caption{Methods of deriving Doppler velocities from the observed spectral-line profiles: (a) Bisector method. 
The dark curve shows the line profile observed at a random location, and the vertical dashed line indicates the wavelength in the rest frame of reference. 
The magenta line shows the calculated bisector of the line, and the $20\%, 40\%, 60\%$, and $80\%$ lines indicate the intensity levels from which \vtwo, \vfour, \vsix, and \veight\ are inferred. 
(b) MDI-like algorithm. 
The Gaussian curves, displayed in different colors, represent the filter sensitivities corresponding to \sdo/HMI's six line positions, and are used to convert the IBIS-observed line profiles into HMI-like observations.
(c) Center-of-gravity method.
The $10\%$ and $90\%$ intensity lines indicate within which intensity levels the center-of-gravity is calculated, and the red star indicates the location of the center of gravity, from which Doppler velocity is calculated relative to the rest frame of reference (the vertical dashed line).}
\label{line_mthd}
\end{figure}

\begin{figure}[!t]
\centering
\includegraphics[width=1.0\textwidth]{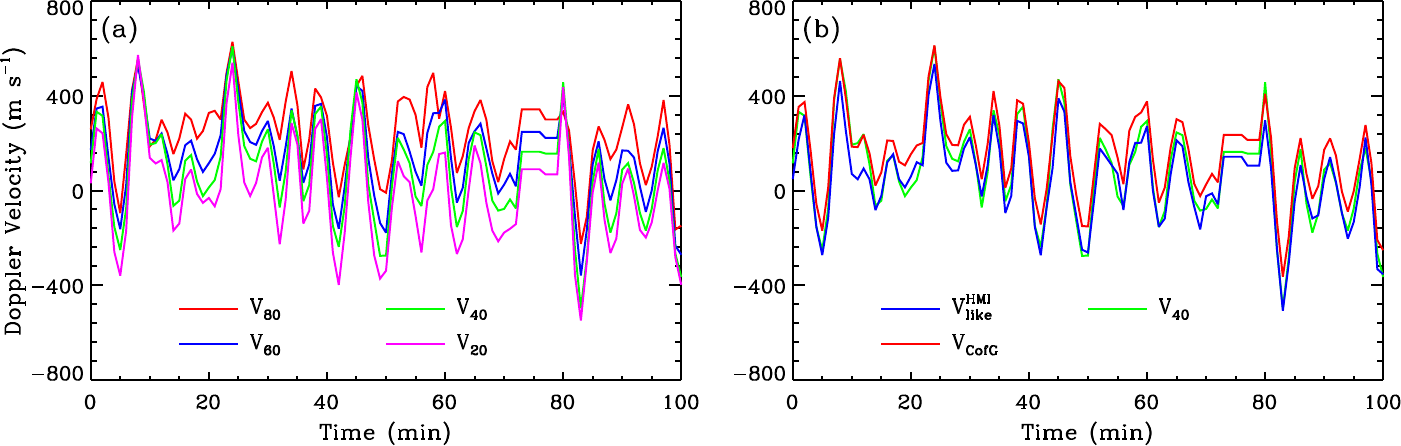}
\caption{(a) Comparison of Doppler velocities derived from different intensity levels using the bisector method, taken at a random location for a random 100-min period. 
(b) Comparison of Doppler velocities \vfour, \vhmi, and \vcog, taken from the same location and time period as the velocities in Panel (a).}
\label{dop_comp}
\end{figure}

In this study, we first derive line-of-sight Doppler velocities from the spectral-line profiles by use of three methods: bisector, MDI-like algorithm, and center-of-gravity method, and then compare velocities from various heights and methods for their relative phase shifts in oscillations. 
For the bisector method, we extract Doppler velocities of plasma motions within the line-forming layers from the line bisectors, in a similar way used in previous studies \citep{raj07, raj10}. 
For the spectral-line profile acquired at each spatial location, we use 10 bisector levels with equal spacing in line intensity to derive 10 Doppler velocities, corresponding to $0\%, 10\%, \ldots, 90\%$ intensity levels and relative to the line-core wavelength 6173.34\,\AA\ in the rest frame of reference (see Figure~\ref{line_mthd}a). 
The Doppler velocities derived from different intensity levels form at different optical depths, which correspond to, approximately, different atmospheric heights, with $90\%$-intensity level slightly above the line's continuum formation height of 100\,km, and $0\%$-intensity level corresponding to line-core formation height at $\sim$270\,km. 
The intensity levels in between are approximately formed evenly between these heights; however, one needs to be cautioned that the line-formation heights vary depending on solar structures.
In this study, for the relative phase shifts measured in Section~\ref{sec3}, we only use Doppler velocities derived for the $80\%, 60\%, 40\%$, and $20\%$-intensity levels, denoted hereafter from the lower to higher atmosphere as \veight, \vsix, \vfour, and \vtwo, respectively. 
Figure~\ref{img}b displays a sample image of \vfour, and Figure~\ref{dop_comp}a shows a comparison of these four velocity curves at a random location with a 100-min duration. 
The comparison (Figure~\ref{dop_comp}a) shows that the amplitudes of the Doppler velocities derived for different atmospheric heights are quite different, larger in higher atmosphere and smaller in lower, but the oscillations at different heights show clear similarities with little visual difference. 

The second Doppler-velocity derivation method, MDI-like algorithm, is the method that was previously used for the data taken by Michelson Doppler Imager onboard {\it Solar and Heliospheric Observatory} \citep[{\it SOHO}/MDI;][]{sch95} and is also currently used in the \sdo/HMI's routine production pipeline of Doppler velocities. 
Because \sdo/HMI only takes observations at 6 line positions while the IBIS observations cover the full line profile, we can determine Doppler velocities in the same way as done for \sdo/HMI.
As shown in Figure~\ref{line_mthd}b, we first multiply the IBIS-observed line profiles with the six \sdo/HMI filter transmission profiles to match \sdo/HMI's 6-point intensities.
We then follow the procedure prescribed by \citet{cou12} to derive the HMI-like Doppler velocities (hereafter, \vhmi):
\begin{equation}
V^\mathrm{HMI}_\mathrm{like} = \frac{\mathrm{d}V}{\mathrm{d}\lambda} \frac{T}{2\pi} \mathrm{atan}\left(\frac{b_1}{a_1}\right), 
\end{equation}
where $\mathrm{d}V/\mathrm{d}\lambda = 48,562.4$\,m\,s$^{-1}$\,\AA$^{-1}$, $T$ is the observed wavelength span equal to $412.8$\,m\AA\ in this case, and $a_1$ and $b_1$ are the first Fourier components:
\begin{equation}
a_1 = \frac{2}{T} \int^{+\frac{T}{2}}_{-\frac{T}{2}} I (\lambda) \cos\left(2\pi\frac{\lambda}{T}\right) \mathrm{d}\lambda,
\end{equation}
\begin{equation}
b_1 = \frac{2}{T} \int^{+\frac{T}{2}}_{-\frac{T}{2}} I (\lambda) \sin\left(2\pi\frac{\lambda}{T}\right) \mathrm{d}\lambda.
\end{equation}

The third Doppler-velocity derivation method, center-of-gravity method (see Figure~\ref{line_mthd}c), computes the Doppler shift corresponding to the intensity-weighted average of the whole line profile following the formula:
\begin{equation}\Delta\lambda = \frac{\sum I_j \lambda_j}{\sum I_j} - \lambda_0,
\end{equation}
where $I_j$ represents the intensity at the $j$-th line position with a wavelength of $\lambda_j$, and $\lambda_0$ is the line-core wavelength in the rest frame of reference.
The center-of-gravity Doppler velocity (\vcog) is then derived from $\Delta\lambda$.

Figure~\ref{dop_comp}b displays a comparison of \vfour, \vhmi, and \vcog\ at a random location for a 100-min duration.
The comparison shows that the Doppler velocities derived from the MDI-like algorithm are similar to \vfour, and the velocities from the center-of-gravity method are similar to the \vsix\ from the bisector method. 
This implies that the atmospheric height corresponding to the \sdo/HMI Doppler velocities is at about the 40\%-intensity level of the \ion{Fe}{1} line.
Despite being obtained from very different methods, the oscillation patterns in these three velocities show remarkable similarities although the following analysis will disclose their small phase differences.

For all the above three Doppler-velocity derivation methods, at each location we first use both the left ($I-V$) and right ($I+V$) circular polarization profiles to derive two Doppler velocities separately, and then average the two as the final velocity to be used in the follow-up analyses.
This is essentially consistent with the standard \sdo/HMI Doppler-velocity derivation procedure, in which two sets of Doppler velocities from the left and right circular polarization are averaged.
After all the Doppler velocities are derived at each location, we then rebin the data by 2$\times$2 in space to enhance the signal-to-noise ratio before carrying out our phase measurements. 

\section{Data Analysis}
\label{sec3}

One main objective of this study is to investigate the relative phase shifts between acoustic waves observed at different atmospheric heights, particularly for the waves with frequencies below the cutoff frequency, i.e., evanescent waves.
It is understandable that in sunspot regions, due to the complicated interaction between helioseismic waves and magnetic field as well as the substantial reduction of the cutoff frequency, the results from sunspot regions will differ significantly from quiet-Sun regions \citep[see][]{raj10}.
Therefore, in this study, we limit our analysis to the quiet-Sun region, delimited by the two dashed circles in Figure~\ref{img}b, and focus only on the phase shifts in evanescent waves between different heights.

\subsection{Cross Spectrum}
\label{sec31}

We first examine the phase-shift diagrams between oscillatory signals observed at different atmospheric heights, i.e., phase shifts in higher atmosphere of \vtwo, \vfour, and \vsix\ relative to the lower atmosphere of \veight. 
The phase-shift diagram as a two-dimensional function of harmonic degree $\ell$ and frequency $\nu$ can be calculated following
\begin{equation}
\delta\phi_\mathrm{n} (\ell, \nu) = \arg \left[ \widehat{V_\mathrm{n}} (\ell_x, \ell_y, \nu) \widehat{V_\mathrm{80}}^\dag (\ell_x, \ell_y, \nu) \right],
\end{equation}
where $\widehat{V_\mathrm{80}} (\ell_x, \ell_y, \nu)$ represents three-dimensional Fourier transform of $V_\mathrm{80}(x,y,t)$, $\ell_x$ and $\ell_y$ are horizontal components of $\ell$ ($\ell = \sqrt{\ell_x^2 + \ell_y^2}$), $\dag$ represents the conjugate of the Fourier transform, $\arg$ represents the argument of a complex numbers, i.e., relative phase, and $\mathrm{n}$ represents one of the three numbers: 20, 40, and 60.

\begin{figure}[!t]
\centering
\includegraphics[width=1.0\textwidth]{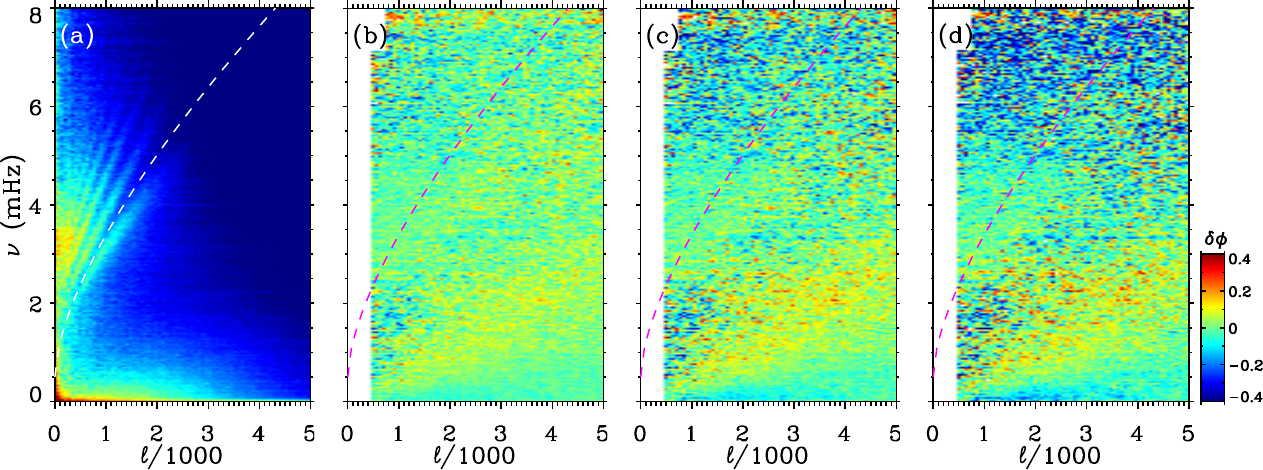}
\caption{(a) Power diagram of the cross spectrum between \vtwo\ and \veight, computed using the entire field of view (including the sunspot) shown in Figure~\ref{img}. 
The power diagrams for other cross-spectra, calculated between other pairs of Doppler velocities, are very similar, thus not shown. 
(b) -- (d)  Phase diagrams of the cross-spectra between \vsix\ and \veight, \vfour\ and \veight, and \vtwo\ and \veight, respectively. 
The phase diagrams are computed using only the quiet-Sun region shown in Figure~\ref{img}, and this explains why the area of $\ell < 500$ is not shown.
The color bar on the right represents the relative phases, with positive (negative) values indicating oscillations in the higher atmosphere leading (lagging behind) those in \veight.
The dashed curves in all panels represent the line separating the $f$- and $p_1$-ridges, below which signals are not used in the phase-shift analyses that follow.}
\label{x_spect}
\end{figure}

Figure~\ref{x_spect}a shows the power diagram of the cross-spectrum between \vtwo\ and \veight, so that the relative locations of acoustic modes (or $p$-modes), $f$-mode, convection, and internal gravity waves can be easily identified. 
Figure~\ref{x_spect}b-d show the phase diagrams of the three higher-atmosphere velocities relative to \veight, calculated using only the quiet-Sun signals.
As can be seen, positive phase shifts (seemingly downward propagating) dominate the areas beneath the $f$-mode ridge, between $\ell$ of $500 - 5000$ and $\nu$ of $0 - 3$\,mHz. 
These are believed to mostly correspond to internal gravity waves, consistent with the results reported by \citet{str08}.
For acoustic modes above the dashed curve, the phase shifts clearly show negative signs (seemingly upward propagating) in higher frequencies and positive signs (seemingly downward propagating) in lower frequencies.
It is not very clear in these plots at what frequency the phase shifts switch the sign from positive to negative. 

\subsection{Relative Phase Shifts between Different Heights}
\label{sec32}

To more accurately measure how phases change in both evanescent and propagating waves above the photosphere, we measure the phase shifts \dphi\ at each spatial location between the Doppler oscillations obtained at higher atmospheric levels and those at \veight\ following the formula: 
\begin{equation}
\delta\phi_\mathrm{n}(x,y,\nu) = \arg \left[ \widehat{V_\mathrm{n}} (x, y, \nu) \widehat{V_\mathrm{80}}^{\dag} (x, y, \nu) \right],
\end{equation}
where $\widehat{V_\mathrm{n}}(x, y, \nu)$ represents the one-dimensional Fourier transform of $V_\mathrm{n}(x, y, t)$, and $\mathrm{n}$ represents one of the three numbers: 20, 40, and 60.
Because we are only interested in studying the behaviors of acoustic waves, all signals corresponding to the $f$-mode, convection, and internal gravity waves are filtered out, and only signals above the dashed curves in Figure~\ref{x_spect} are kept.
Then a two-dimensional Gaussian smoothing is applied on the $x$-$y$ space to further enhance the signal-to-noise ratio.
The FWHM of the Gaussian function is $1\farcs5$ for the results presented in this article; however, different values of FWHM are tested and the analysis results remain largely unchanged. 
After the data are filtered and smoothed, for each of the three higher atmospheric levels, the $\delta\phi_\mathrm{n}(x,y,\nu)$ is measured at each quiet-Sun location, and then all the $\delta\phi\mathrm(x,y,\nu)$ are averaged in $x$-$y$ space for $\delta\phi\mathrm(\nu)$.

\begin{figure}[!t]
\centering
\includegraphics[width=1.0\textwidth]{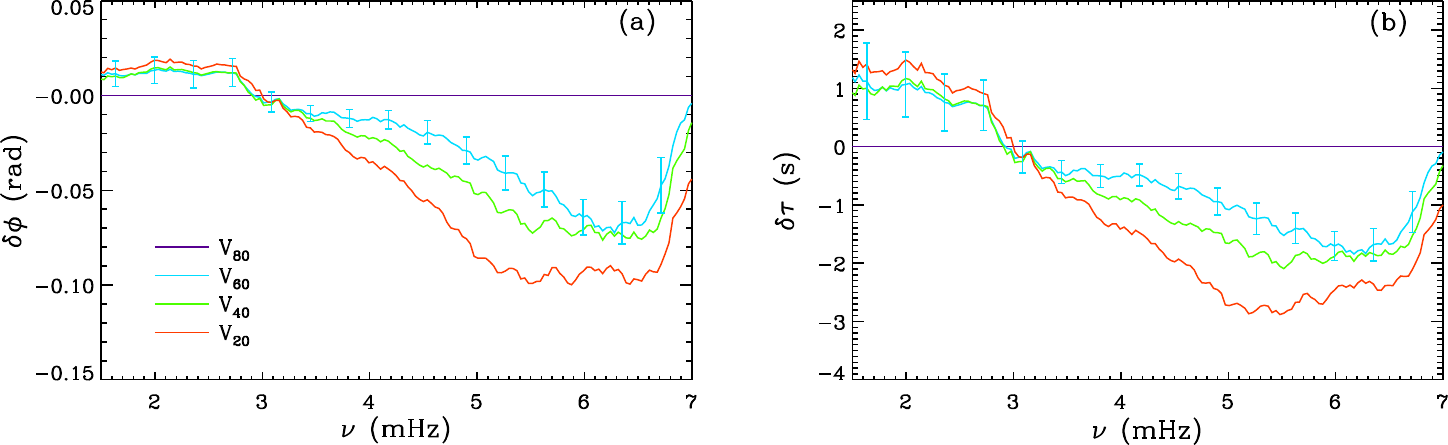}
\caption{(a) Phase shifts \dphi\ measured from \vtwo, \vfour, \vsix, and \veight\ relative to \veight, displayed as functions of frequency $\nu$. 
For clarity, standard errors are displayed only for selected points in one of these curves, and the errors for other points in this curve and in other curves are similar.
(b) Same as Panel (a) but relative time shifts \dt\ are displayed. }
\label{phase_rel2}
\end{figure}

Figure~\ref{phase_rel2} shows the relative phase shifts \dphi\ and travel-time shifts \dt\ measured for all atmospheric levels as functions of $\nu$, averaged from the entire quiet-Sun region where the net vertical flow is presumed to be close to 0.
The \dt\ is computed from \dphi\ following $\delta\tau(\nu) = \delta\phi(\nu)/2\pi\nu$.
It can be seen that the \dphi, albeit very small between 1.5 and 4.0\,mHz, show positive values below $\sim$3.0\,mHz and negative values above $\sim$3.0\,mHz, indicating that even for evanescent waves in a region with no or little net vertical flows, the measured phases continue to change with height. 
The positive values indicate that the phases in the higher atmosphere lead those in the lower atmosphere, as if the waves propagate downward.
Negative phase shifts persist from 3.0\,mHz to 7.0\,mHz and continue to grow in values with frequency. 
It is not surprising for the phases of $\nu >$ 5.0\,mHz waves to grow because these waves are generally believed to be propagating waves, but it is a bit surprising that the magnitude of \dphi\ drops beyond $\sim$6.0\,mHz.
This may be due to the increased observational noises with the increase of acoustic frequency, or due to the acoustic halos \citep[e.g.,][]{raj13, rij16} outside of the sunspot where the data are taken for these analyses.
However, whether the measured values match the expected values above 5.0\,mHz and how to explain the phase behaviors above 6.0\,mHz are beyond the scope of this article.
Figure~\ref{phase_rel2}b shows that even for the relatively small \dphi's, the corresponding \dt's can be around 1\,s or larger, implying that these small phase changes can cause substantial errors for some local helioseismic measurements, particularly when travel-time shifts caused by very weak flows are measured, such as meridional circulation in the Sun's deep interior or vertical flows beneath supergranules (see discussions in Section~\ref{sec41}).

\begin{figure}[!t]
\centering
\includegraphics[width=0.55\textwidth]{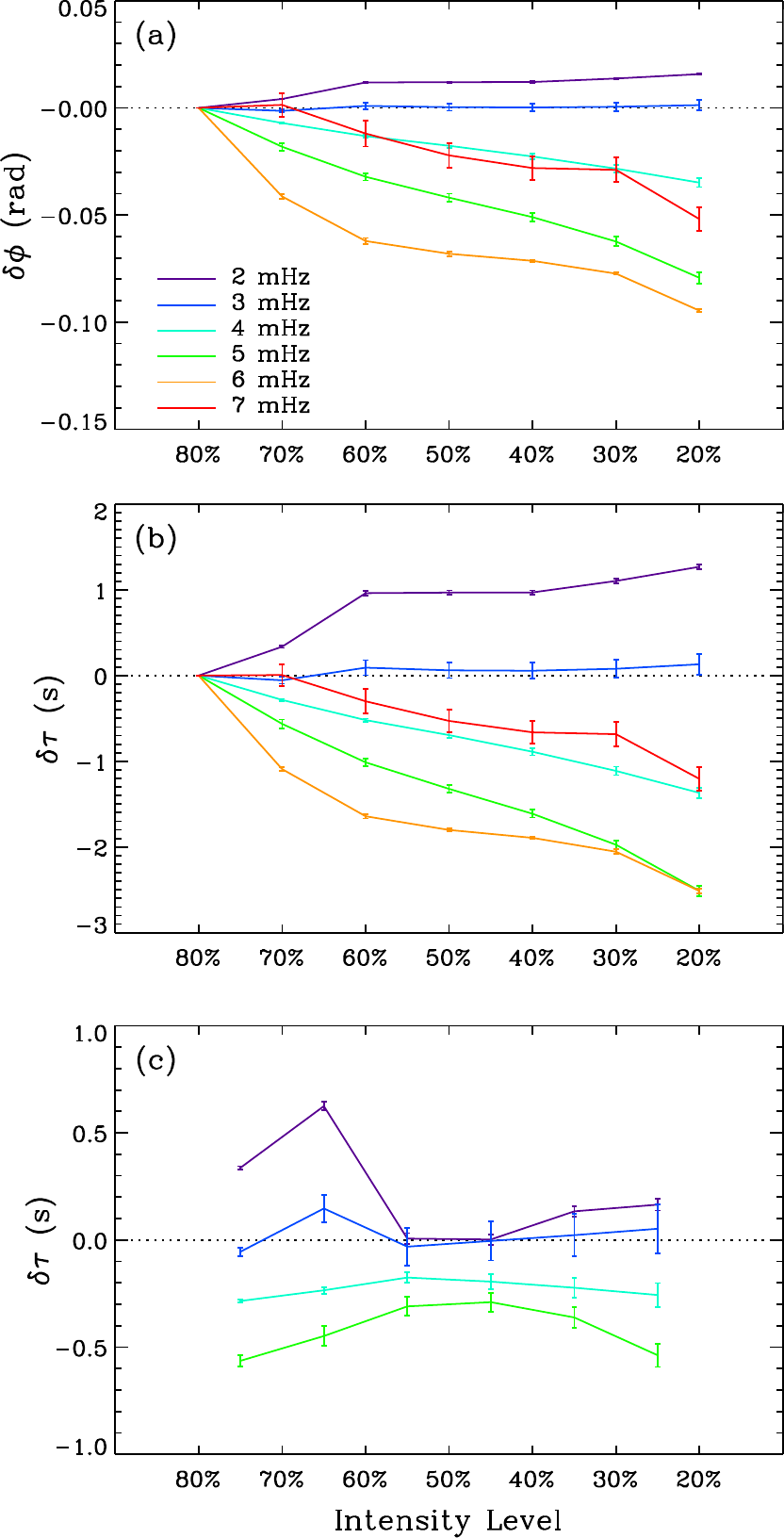}
\caption{(a) Relative phases shifts \dphi, displayed as functions of intensity level for different frequency bands. 
(b) Same as Panel (a) but relative time shifts \dt\ are displayed. 
(c) Differences of \dt\ between one atmospheric level and its higher level, displayed as functions of height for different frequencies.
For clarity, results for frequency bands of 6.0 and 7.0\,mHz are not displayed.}
\label{phase_ht}
\end{figure}

Figure~\ref{phase_rel2} also shows that \dphi\ (\dt, too) are not just functions of $\nu$, but also functions of height: while the positive \dphi\ remain largely unchanged with height, the negative \dphi\ grow substantially. 
Figure~\ref{phase_ht} shows both \dphi\ and \dt\, as well as the change rate of \dt\, as functions of intensity level (or approximately, atmospheric height), obtained for a few 1.0-mHz-wide frequency bands with their middle frequencies marked in the figure. 
In most atmospheric layers, the \dphi\ around 2.0\,mHz remain largely flat and positive, and the corresponding \dt\ are around 1\,s. 
The \dphi\ around 3.0\,mHz also remain mostly flat, close to 0 across all layers. 
For all the other frequency bands above 3.0\,mHz, the \dphi\ are negative and have a trend of growth with height, and the \dt\ values are of an order of 1--2\,s. 
Figure~\ref{phase_ht}c shows the phase changing rate, i.e., how rapid phase changes with height, is mostly similar for a same frequency in different heights, except a few notable points.

\begin{figure}[!t]
\centering
\includegraphics[width=1.0\textwidth]{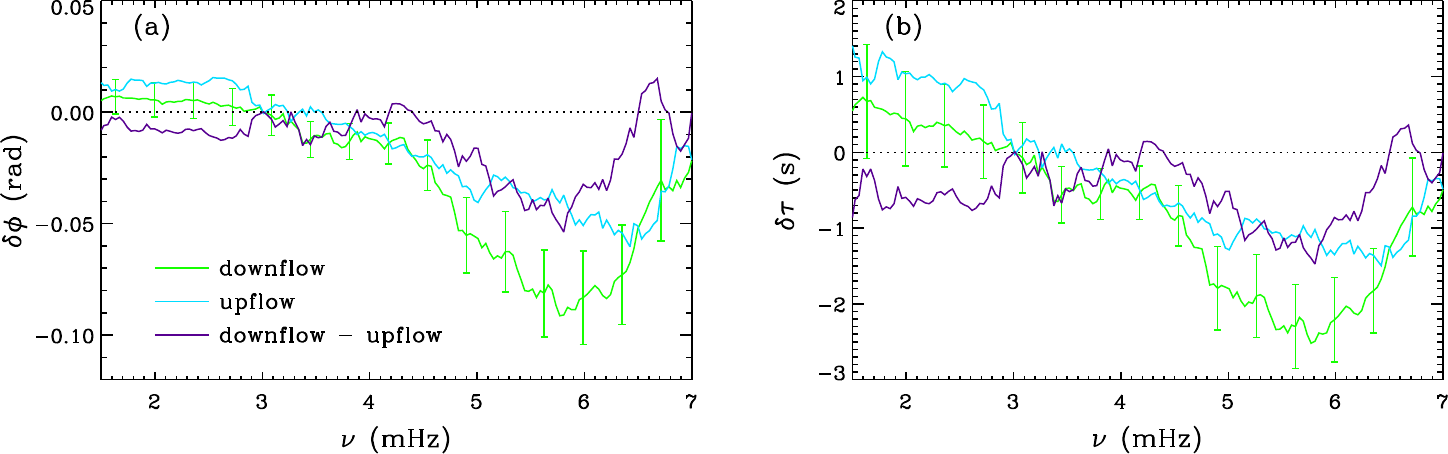}
\caption{(a) Relative phase shifts \dphi\ measured between \vsix\ and \veight\ inside the regions with strong downward and upward flows, as well as the differences between the two \dphi\ curves.
For clarity, error bars are only displayed in selected points in one curve. 
The errors are similar for other points of this curve and for other curves. 
(b) Same as Panel (a) but relative time shifts \dt\ are displayed. }
\label{phase_updown}
\end{figure}

To investigate behaviors of acoustic waves in areas with persistent upward and downward flows, we select areas with the top 10\% of upflow speeds and the top 10\% of downflow speeds for further analysis. 
After removing a uniform background velocity, i.e., caused by rotation, the selected upflow (downflow) region has an average speed of $-190$ ($+190$) \mps\ during the observation period. 
Figure~\ref{phase_updown} shows that the \dphi\ and \dt, measured from \vsix\ relative to \veight, exhibit a frequency-dependent variation trend similar to those measured in the entire quiet region (Figure~\ref{phase_rel2}). 
It is also clear that the \dt\ measured in the upflow region is systematically above the \dt\ measured in the downflow region, with the differences around 0.5\,s. 
Note that this difference between the two \dt\ can be seemingly explained by the upward (downward) flows that speed up (slow down) both evanescent and propagating waves, but as a matter of fact, the measured values are too large to be explained this way. 
Given that the sound speed is much greater than 10\,\kmps\ above the photosphere and the total distance between these two layers is $<$50\,km, a flow speed difference of 380\,\mps\ cannot cause a \dt\ that matches the measured values. 
Measurements between \veight\ and other higher atmospheres give similar values below 3.0\,mHz and considerably larger values above that.

\subsection{Relative Phase Shifts between Doppler Velocities from Different Methods}
\label{sec33}

\begin{figure}[!t]
\centering
\includegraphics[width=1.0\textwidth]{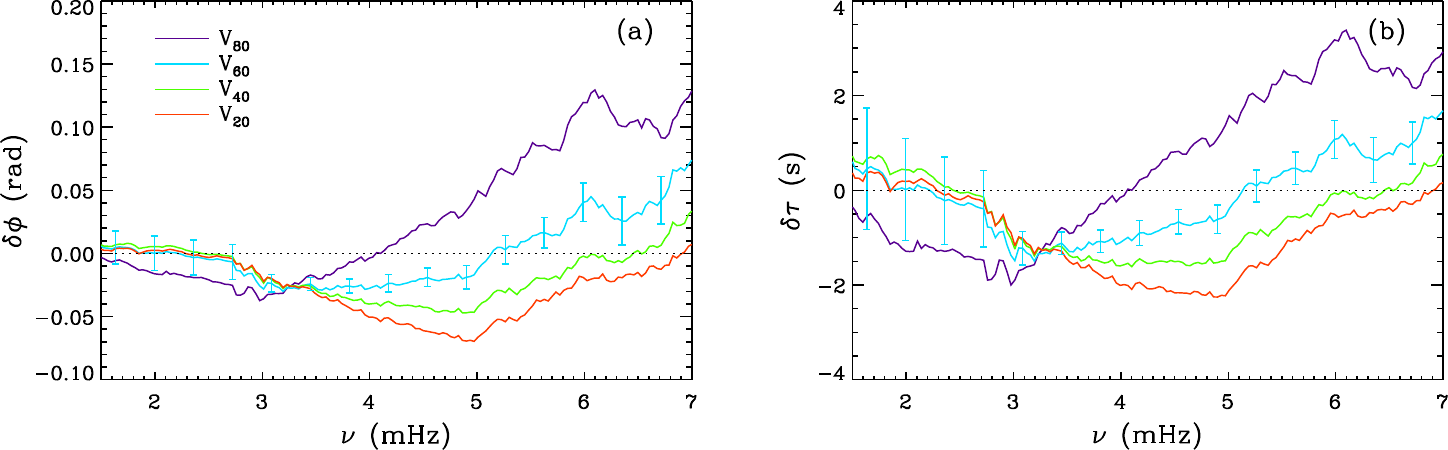}
\caption{(a) Phase shifts \dphi\ measured from \veight, \vsix, \vfour, and \vtwo\ relative to \vhmi, displayed as a function of frequency $\nu$. 
For clarity, standard errors are displayed only for selected points in one of these curves, and the errors for other points of this curve and for other curves are similar.
(b) Same as Panel (a) but relative time shifts \dt\ are displayed. }
\label{rel_hmi}
\end{figure}

As noted in Section~\ref{sec2}, in addition to the bisector method, the MDI-like algorithm and the center-of-gravity method are also used to derive Doppler velocities from the observed spectral-line profiles. 
It is also of interest to examine relative phase shifts in the Doppler velocities derived from these different methods. 
Figure~\ref{rel_hmi} shows \dphi\ of the four bisector Doppler velocities relative to \vhmi\ as functions of $\nu$.
It can be found that the phases of \vhmi\ are relatively close to the phases of the bisector velocities below $\sim$3.0\,mHz, but beyond that their phases show substantial differences. 
The \dphi\ of the bisector Doppler velocities relative to the \vcog\ (plots not shown in this article) give similar results.

\begin{figure}[!t]
\centering
\includegraphics[width=1.0\textwidth]{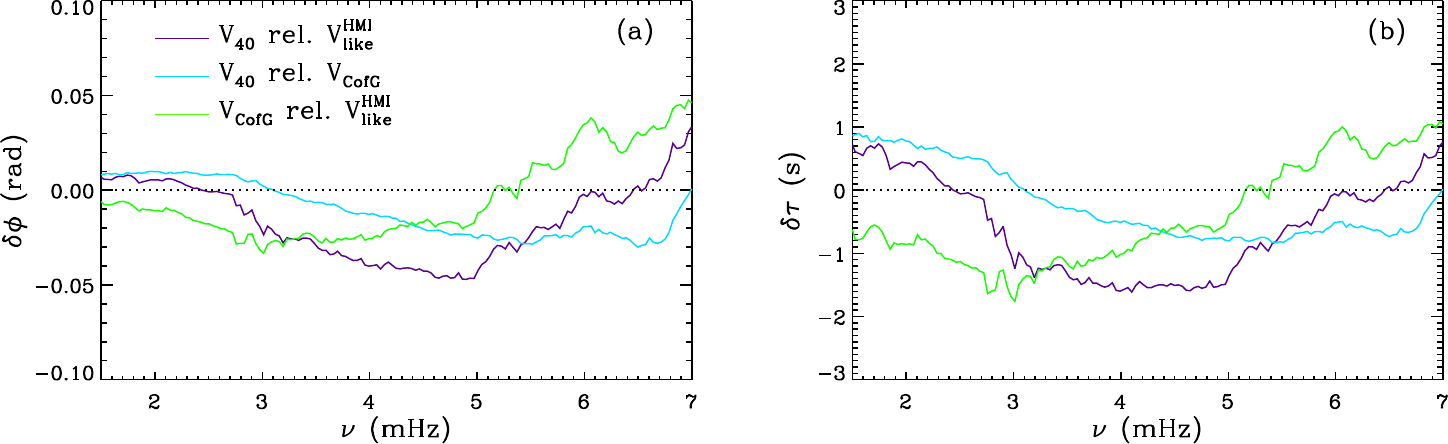}
\caption{(a) Phase shifts \dphi\ measured from \vfour\ relative to \vhmi, \vfour\ relative to \vcog, and \vcog\ relative to \vhmi. 
Error bars are similar to those shown in Figure~\ref{rel_hmi}a thus not plotted here. 
(b) Same as Panel (a) but relative time shifts \dt\ are displayed. }
\label{mdi_cofg_40}
\end{figure}

As demonstrated in Figure~\ref{dop_comp}b, both the \vhmi\ and \vcog\ show similar magnitudes and variation patterns to \vfour, indicating that these two Doppler velocities correspond best with the bisector velocity inferred for the 40\%-intensity level. 
Figure~\ref{mdi_cofg_40} shows \dphi\ and \dt\ between these three Doppler velocities. 
None of the pairs fully agree with each other in the frequency-dependent \dphi, indicating that all the Doppler-deriving methods show a different amount of combination of information from various atmospheric heights, and these combinations seem to be frequency dependent as well. 

\subsection{Comparing IBIS and \sdo/HMI Results}
\label{sec34}

We have by now examined the relative phase shifts in Doppler velocities obtained at different atmospheric heights and from different velocity-derivation methods. 
It would be interesting to examine how the oscillatory phases observed in intensities change with height, and how the measured results from IBIS observations compare with those from \sdo/HMI. 
Unfortunately, for both IBIS and \sdo/HMI, intensities can only be reliably inferred at line core and continuum of the spectral line but not for other optical depths; IBIS data were acquired long before \sdo\ was launched, thus simultaneous comparison of IBIS and \sdo/HMI data is not possible; and, \sdo/HMI does not have multiple Doppler velocities available like those used in Sections~\ref{sec32} and \ref{sec33}. 
Double-height Doppler-velocity proxies were attempted previously using \sdo/HMI's 6-position data \citep{nag14}, but the accuracy of those data cannot meet the high-precision requirements of this study.
Therefore, in this study for both IBIS data and one selected dataset from \sdo/HMI, we compute relative phase shifts \dphiI\ between line-core intensity (\ilc) and continuum intensity (\ic), which are then compared.
For both sets of data, we also compute the relative phase shifts, \dphiVL\ and \dphiVI, between Doppler velocities $V$, derived using the MDI-like algorithm for both IBIS and \sdo/HMI data, and their corresponding \ilc\ and \ic, respectively. 
However, meanwhile, one needs to keep in mind that the velocity and intensity data show different senses of mode asymmetries \citep{duv93a}, and discussions related to this are in Section~\ref{sec43}.

\begin{figure}[!t]
\centering
\includegraphics[width=1.0\textwidth]{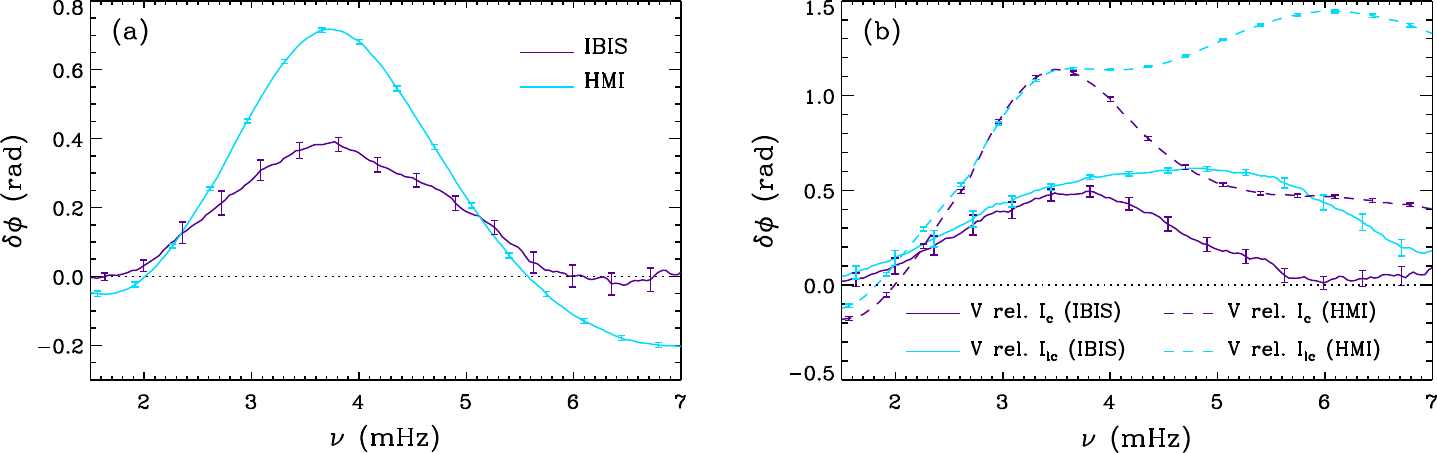}
\caption{(a) Phase shifts measured from line-core intensities relative to continuum intensities, for both the IBIS and \sdo/HMI observations. 
(b) Phase shifts measured from Doppler velocities, which are derived using MDI-like algorithm, relative to line-core and continuum intensities, respectively, for both the IBIS and \sdo/HMI observations.
}
\label{HMI_IBIS}
\end{figure}

In this study, \ic, \ilc, and $V$ data are taken from 2011 January 1 \sdo/HMI observations near the disk center, with a duration of 464\,min (same as the IBIS data duration) and covering an area of $256\arcsec \times 256\arcsec$ with a 45-sec cadence. 
The \sdo/HMI spatial resolution is $1.0\arcsec$ (or $0.5\arcsec$ pixel$^{-1}$), much coarser than the IBIS data; but the field of view of this selected dataset is nearly as 10 times large as the IBIS data. 
Figure~\ref{HMI_IBIS}a shows \dphiI\ for both datasets, in which it can be seen that the trends of the curves are similar, with oscillations in the higher atmosphere lead those in the lower atmosphere substantially between about $2.0 - 5.5$\,mHz, although the sign-reversal frequency differs in these two curves. 
However, it is also clear that the \dphiI\ values from \sdo/HMI are substantially larger than (about twice of) those from IBIS.

Figure~\ref{HMI_IBIS}b shows \dphiVL\ and \dphiVI\ for both sets of observations, in which $V$ leads both \ic\ and \ilc\ substantially in phases. 
Again, the variation trends for both \dphiVL\ and \dphiVI\ curves show many similarities in the IBIS and \sdo/HMI results; however, the \dphi\ measured from the \sdo/HMI data are about as twice large as those measured from the IBIS data. 

\begin{figure}[!t]
\centering
\includegraphics[width=1.0\textwidth]{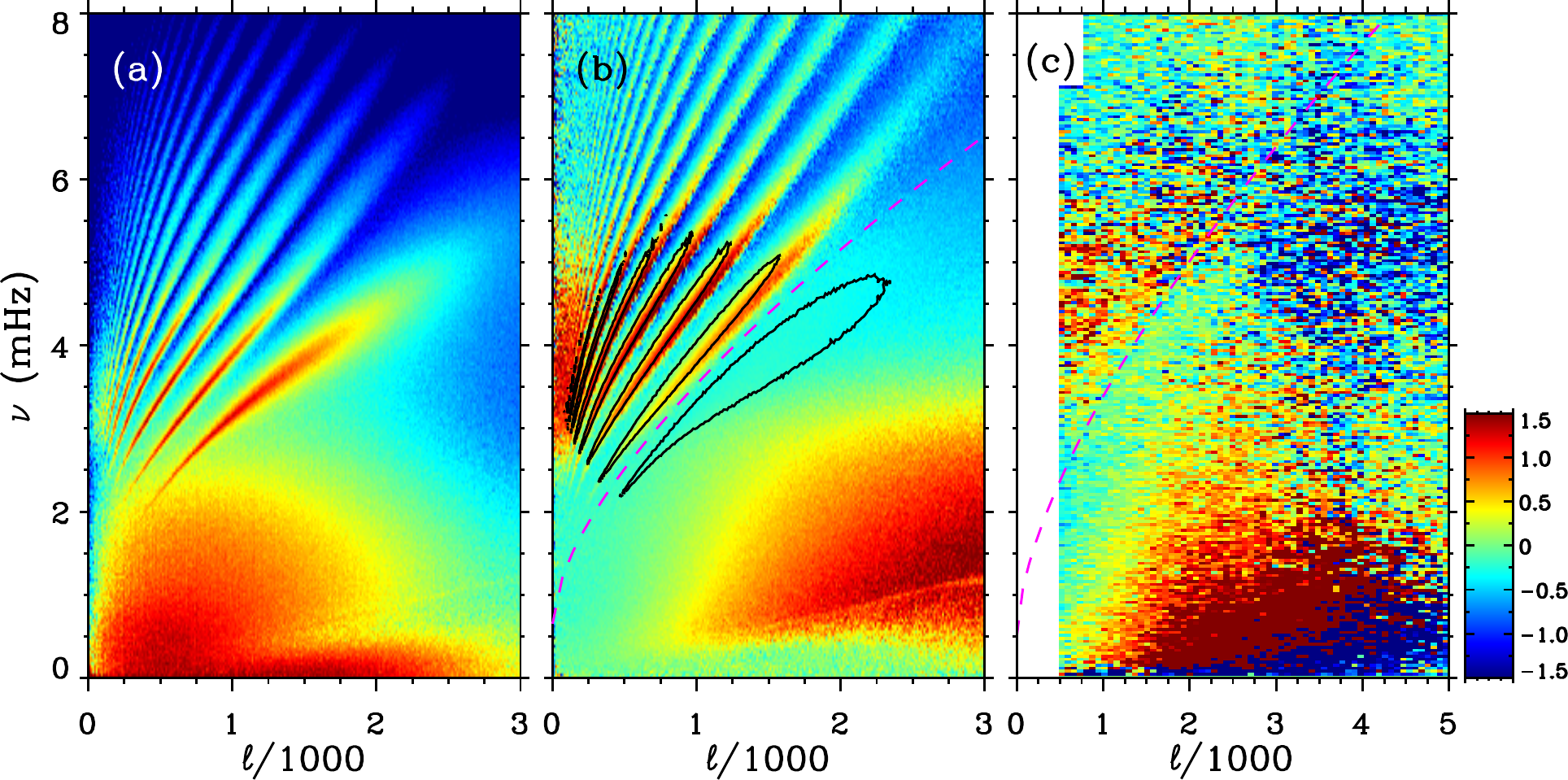}
\caption{(a) Power diagram for the cross spectrum between the \sdo/HMI-observed line-core intensity and continuum intensity.
(b) Phase diagram for the same cross spectrum, with the logarithm of the power overplotted as contours to show relative locations of the power ridges.
(c) Phase diagram for the cross spectrum between the IBIS-observed line-core intensity and continuum intensity. 
Same as Figure~\ref{x_spect}b-d, the low-$\ell$ area is left blank in the plot.
The color bar on the right shows color scales of relative phases in both panels (b) and (c). 
The magenta dashed lines in panels (b) and (c) indicate the middle line separating $f$ and $p_1$ ridges, below which signals are filtered out in the phase-shift analyses presented in Section~\ref{sec34}.}
\label{xspect2}
\end{figure}

To understand the differences in the IBIS and \sdo/HMI results, we compute phase diagrams of the cross-spectra between \ic\ and \ilc\ for both datasets (Figure~\ref{xspect2}b and \ref{xspect2}c). 
Understandably, because of its higher spatial resolution and smaller field-of-view, the IBIS data has poorer wavenumber (or harmonic degree $\ell$) resolution relative to the HMI data. 
Both of the phase diagrams show similar and clear frequency dependency that is also seen in Figure~\ref{HMI_IBIS}: large positive \dphi\ in the range of about $2.0 - 5.5$\,mHz and small negative \dphi\ above about 5.5\,mHz. 
Figure~\ref{xspect2}b also shows at least two \dphi-variation trends: 
(1) The \dphi\ show alternating positive and negative signs on either side of the $p$-mode power ridges, with positive values overlapping more with the power ridges, which results in giving a net positive sign in phase-shift analysis shown in Figure~\ref{HMI_IBIS}a; 
(2) Despite the alternating signs of \dphi, there is a general trend of decreasing positive values and increasing negative values with the increase of $\ell$. 
The trend (2) explains, we tend to believe, why the values of \dphi\ measured from IBIS and \sdo/HMI differ substantially (Figure~\ref{HMI_IBIS}): the IBIS data are more sensitive to the higher-$\ell$ wave signals, getting more contributions from the low-\dphi\ portion to the measured \dphi\ than the \sdo/HMI data.

\section{Discussion}
\label{sec4} 

\subsection{Impacts to Local Helioseismology Studies}
\label{sec41}

In this study, we have analyzed relative phase shifts, \dphi, between oscillatory signals observed in Doppler velocities of different optical depths (approximately, different atmospheric heights), derived using a set of long-duration IBIS observations with full spectral-line profiles, high spatial and spectral resolutions, and a high temporal cadence. 
Due to the scarcity of high-quality simultaneous Doppler observations of multiple atmospheric heights, similar types of analysis were rare in the past.
We believe that the \dphi, as a function of frequency and atmospheric height, reported in this article have profound impacts to local helioseismology studies, including time--distance helioseismology \citep{duv93} among others.

As introduced in Section~\ref{sec1}, the Sun's acoustic cutoff frequency (\nuac) in the photosphere is around 5.0\,mHz \citep[e.g.,][]{deu84, jim11}, thus our observed waves consist of both evanescent and propagating waves.
It is often presumed that the phases of evanescent waves remain unchanged with height unless the waves are perturbed by vertical flows or other dynamic parameters; or if perturbed, local helioseismic inversions are expected capable of recovering these perturbations.
However, our analysis shows that for the acoustic-mode oscillations inferred from the \FeI\ line in a quiet-Sun region, where the net vertical flow is expected to be close to 0 (for discussions of the role of convective blueshift, please refer to Section~\ref{sec43}): relative to lower-atmosphere oscillations, higher-atmosphere oscillations lead in phases while $\nu$\,$<$\,3.0\,mHz and lag behind while $\nu$\,$>$\,3.0\,mHz (for discussions of evanescent waves, please refer to Section~\ref{sec42}). 
These measured \dphi\ are not consistent with the general presumption of the stationary phases in evanescent waves, nor can be simply explained by vertical flows. 
The long-term (around 7\,hrs in this study) average of the vertical flows, if not exactly net 0, is not expected to exceed an order of 100\,\mps, which, to the waves with a phase speed $>$10\,\kmps\ within a travel interval of $<$200\,km, can hardly cause phase (or travel-time) shifts that match the measured values in Figures~\ref{phase_rel2} \& \ref{phase_ht}.
Equally importantly, vertical flows cannot explain the measured frequency dependency of \dt, either. 

\begin{figure}[!t]
\centering
\includegraphics[width=0.6\textwidth]{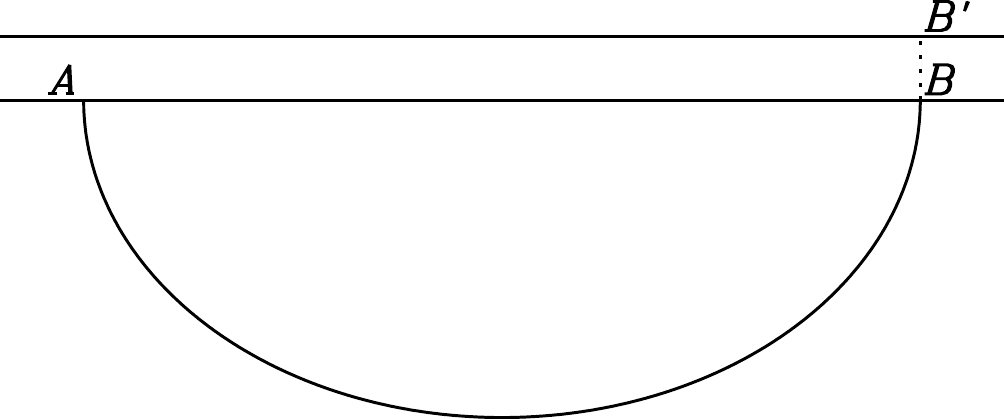}
\caption{Schematic plot showing a subsurface ray path connecting two surface locations $A$ and $B$, along which waves travel between these two locations.
Signals presumably observed at location $B$ are actually observed at a slightly higher location $B^\prime$.
Plots are not to scale.}
\label{schematic}
\end{figure}

What possibly causes these observed phase shifts will be discussed in Section~\ref{sec43}, and here we focus on how these results impact local helioseismic studies. 
Let us take time--distance helioseismology as example, which measures (see Figure~\ref{schematic}) travel time \tab\ of acoustic waves traveling from one surface location $A$ to another surface location $B$ through the solar interior, and measures travel time \tba\ along the opposite direction. 
Suppose the Doppler observation at location $A$ forms at one atmospheric level, and the Doppler observation at location $B$ forms at a slightly higher level --- location $B^{\prime}$. 
Based on our earlier analysis, there is a travel-time shift \tbbp\ between positions $B$ and $B^{\prime}$; therefore, the presumed \tab\ measurement becomes \tabp, which is equal to \tab +  \tbbp, and the presumed \tba\ measurement now becomes \tbpa, which is equal to \tba $-$ \tbbp. 
The travel-time difference, \dtab $\equiv$ \tab $-$ \tba, which is often interpreted as caused by the plasma flows along the subsurface travel path, becomes \dtab\, + 2\tbbp. 
A measurement error of 2\tbbp\ is thus introduced, which will be mistakenly interpreted as interior flows in the inversions.
Based on our analysis shown in Section~\ref{sec32}, this \tbbp\ can be around $+1$\,s between 1.5 -- 3.0\,mHz and around $-1$\,s between 3.0 -- 5.5\,mHz, approximately the same order of magnitude as caused by the near-surface meridional flow, thus non-negligible.
Actually, this model is not strictly accurate because oscillations at location $B$ and $B^\prime$ consist of all types of wave modes, while waves traveling from $A$ to $B$ consist of only a few modes. 
So the \tbbp\ measured locally may not be identical to the \tbbp\ in the waves traveling to here from $A$.
Nevertheless, it is still reasonable to believe, although the values may slightly differ, using oscillatory signals observed at different atmospheric heights to measure \dphi\ (or \dt) can introduce non-negligible errors.

As a matter of fact, it is not rare that observed oscillatory signals form at different atmospheric heights on and above the Sun's photosphere, albeit most times with only small height differences. 
For instance, due to the limb-darkening effect, the formation heights of most spectral lines gradually increase with the distance to the disk center; and in fact, this is believed to be directly linked to the helioseismic center-to-limb effect \citep{che18}. 
Due to Wilson Depression, the observed sunspot umbra, penumbra, and quiet-Sun regions can differ in heights by $\sim$500\,km, likely causing a few seconds of \dt\ measured between them, which are not due to any flows but were mistakenly ascribed to flows in some past helioseismic inversions \citep[e.g.,][]{zha03}.
Another example is supergranules, which are often thought of as quiet-Sun regions but their central areas may be slightly warmer, thus spectral lines there form slightly higher, than their boundary areas. 
For a long-time (a few hours) analysis typically required by time--distance helioseismology, the \dt\ measured between supergranular centers and boundaries will likely carry a systematic shift that is unrelated to the supergranules' subsurface structure or flows, but such a systematic shift is not accounted for in the time--distance analysis \citep[e.g.,][]{duv13}.

Another important implication to the time--distance helioseismic measurements is in the areas with systematic long-term (an order of a few hours) up- or down-flows. 
As illustrated in Figure~\ref{phase_updown}, the measured travel-time shifts in the areas with up- or down-flows cannot be accounted for by the vertical flows inferred from Doppler shifts, and some other factors must play a role in causing these measured shifts. 
Therefore, if one end of the acoustic waves used for time--distance measurements (say, location $A$ or $B$ in Figure~\ref{schematic}) is inside a region with systematic vertical flows, one needs to be cautious in interpreting the measured \dt. 
There is no shortage of such locations on the solar surface, e.g., again, the boundaries (central areas) of supergranules are usually associated with downward (upward) flows.
This vertical-flow-related effect may either be just another manifestation of the effect caused by the different line-formation heights that is discussed above, because supergranular upflow (downflow) areas are often associated with hotter (cooler) areas, or it is an independent effect in addition to the line-formation height.

It is now clear that the unaccounted-for phase shifts in evanescent waves impact significantly how we interpret local helioseismic measurements, but it is not immediately clear how this observational fact impacts global helioseismology.
Intuitively, the systematic change of oscillatory phases with distance to the disk center, i.e., the center-to-limb effect, albeit small, can systematically change the ridge location $(\ell,\nu)$ in the power spectrum, hence change the inversion for sound-speed structures inside the Sun. 
The effect on the inference of the internal rotation by global helioseismology should be negligible, because the effect from both sides of the central meridian can cancel each other.
In addition, our observations can only identify non-zero phase shifts in evanescent waves in and above the photosphere, but unable to investigate how phases change beneath the photosphere.
For example, let us consider 3.0\,mHz waves, which become evanescent a few hundred km beneath the photosphere: are extra phase shifts introduced before they are observed at the surface? 
A tiny change of their phases in this depth interval that are unrelated to flows will greatly challenge our interpretation of both global and local helioseismic measurements.

Our analysis also shows that the Doppler velocities derived from the MDI-like algorithm and the center-of-gravity method both show small yet non-negligible phase shifts relative to the velocities that are derived from the bisector method and match those two methods best.
This does not imply that any of these Doppler-velocity derivation methods introduce extra phase shifts, but caution us that it may not be appropriate to measure relative phase shifts between velocities derived using different methods. 

\subsection{Evanescent Waves}
\label{sec42}

Our analysis on this specific set of IBIS observations shows that for most atmospheric heights, the \dphi\ of the acoustic waves switches from positive to negative at around 3.0\,mHz, where a mixture of both evanescent and propagating waves is expected. 
However, what frequency separates evanescent waves from propagating waves is not very clear in this set of observations.

Studying the helioseismic center-to-limb effect as a function of acoustic frequency, \citet{che18} pointed out that the sign reversal of the center-to-limb effect occurs at about 5.5\,mHz, which is close to the cutoff frequency, in most regions on the solar disk, and this sign-reversal frequency decreases with distance to the disk center. 
If the center-to-limb effect is indeed, most likely it is, connected to the spectral-line's formation height, we would expect the observed sign reversal in our present analysis to occur around 5.5\,mHz instead of near 3.0\,mHz, because the analyzed region is very close to the disk center and both this analysis and their analysis \citep{che18} used observations from the same spectral line \FeI. 
A few factors may play a role in causing this apparent discrepancy: 
first, in this study the IBIS-observed region is in the vicinity of a sunspot, with visible pores in our ``quiet-Sun" region. 
The presence of magnetic regions, though sparse, and the systematic moat flows around the sunspot, will undoubtedly affect our results while the magnitude of the influence is difficult to assess. 
Second, the sign reversal near 3.0\,mHz seen in this analysis is from the bisector-derived Doppler data, and it is difficult to estimate how the MDI-like algorithm-derived Doppler velocities would behave as a function of frequency and height, thus it is difficult to compare these results directly against the \sdo/HMI results. 
Third, it is possible that different instruments with different spatial resolutions and modulation transfer functions, despite using the same spectral line, can cause systematic differences in the measured results.
All these above factors are for Doppler velocities, but the results obtained from intensity data, presented in Section~\ref{sec34}, may help shed light on understanding more of the sign-reversal frequency.

The comparison of measurements, made using IBIS and \sdo/HMI line-core and continuum intensities, show similar frequency-dependent trends of \dphi\ but with substantially different values.
The phase diagrams of the cross-spectra between the two intensities from both datasets help to explain the discrepancies in the \dphi\ values: the \dphi\ are not just a function of frequency and height, but also a function of wavenumber. 
With a higher spatial resolution, the IBIS \dphi\ measurements are constituted more of low phase shifts than the \sdo/HMI observations, causing the discrepancies in the measured values. 
Although this conclusion is only for intensity data, we speculate the similar thing for the Doppler data. It would be reasonable to speculate that the \dphi\ in \sdo/HMI-observed Doppler data will show a frequency- and height-dependency similar to those measured from IBIS data, but likely with higher values and a higher sign-reversal frequency.


Overall, although it is unclear at what frequency the acoustic waves become evanescent and whether the phase-shift sign reversal is directly related to the cutoff frequency, it is quite clear that the phases of the evanescent waves continue to change, albeit rather small, with the atmospheric height. 
The sign and amount of the phase shifts seem to depend upon acoustic frequency, as well as spectral lines, telescopes, and instruments that are used for observations in addition to vertical flows.

\subsection{Physical Causes of Phase Shifts}
\label{sec43}

Our analysis has shown that above the Sun's photosphere, the measured phases of evanescent acoustic waves continue to shift, and this frequency-dependent shift cannot be explained by vertical flows in the region. What causes such phase shifts?

For an enclosed ideal and adiabatic gas that is experiencing a periodic variation, the velocity change is expected to lead the temperature change in phase; however, the Sun's atmosphere is neither enclosed nor adiabatic. 
For an evanescent wave, even if the phases of oscillatory signals in velocities do not change or only change subtly with atmospheric height, the temperature responses to them, thus the observed intensity changes, at different heights may not be simultaneous, but likely with a gradual height-dependent time delay due to the non-adiabaticity of the atmosphere. 
That is, intensities at different heights do not respond simultaneously to a same velocity perturbation.
As clearly shown in Figure~\ref{phase_rel2}a and Figure~\ref{xspect2}a, oscillatory signals show small phase changes with height when observed in Doppler velocities, but show substantial, an order of magnitude larger, phase changes with height when observed in intensities.
Now, we need to remind ourselves that all Doppler velocities are inferred from spectral-line intensities, and a small leakage from the large height- and frequency-dependent phase shifts in intensities can result in non-negligible phase shifts in Doppler velocities, even when local velocities do not carry such phase shifts intrinsically.
Indeed, \FeI\ line is not perfectly symmetric, and its asymmetry gets enhanced by convective blueshifts \citep[e.g.,][]{loh18, sti19}. 
And, it is also well known that the red wing of spectral lines oscillates with larger amplitudes, thus carrying stronger oscillatory power, than the blue wing \citep{cav85, ber90}.
Therefore, the Doppler velocities derived from this asymmetric line profile, with unbalanced oscillatory power in both of its wings, may carry a height-dependent phase shift leaked in from intensities. 
That is, a combination of the atmosphere's non-adiabaticity and the spectral line's properties can cause height- and frequency-dependent phase shifts in Doppler velocities. 
This is a scenario that we speculate can explain most of the observed phase shifts while the rest can be explained by the scenarios discussed below; meanwhile, we also recognize that this scenario needs to be confirmed using numerical simulations \citep[e.g.,][]{kit15}, and perhaps, more disk-location-dependent observations with the full spectral-line profiles are needed for a more systematic analysis.

Other factors may also play a role in the measured phase shifts. 
To explain the helioseismic center-to-limb effect, which is probably a manifestation of the frequency- and height-dependent \dphi\ reported in this work, \citet{bal12} tried to explain that the ubiquitous convective blueshift above the photosphere likely caused negative phase shifts in both the evanescent and propagating waves. 
(Note that the role of convective blueshift here is to shift phases of acoustic waves, different from its role in causing the spectral-line asymmetry mentioned above.)
This effect undoubtedly plays a role in causing some phase shifts, but the amount of the shifts estimated from numerical models are about one order of magnitude smaller than the observed values, indicating that other factors may play a leading role.
In addition, this mechanism cannot explain the observed frequency dependence of the phase shifts.

Another cause of the observed \dphi\ may be related to the line-asymmetry \citep{duv93a} across the oscillation power ridges. 
(Note that the line asymmetry here is across acoustic power ridges, different from the optical spectral-line asymmetry mentioned above.)
It appears (see Figure~\ref{xspect2}b) that the \dphi\ observed in intensities show opposite signs on either side of the power spectrum ridges, and the line asymmetry can cause unequal contributions to the \dphi\ in our \dphi\ or \dt\ measurements.
If the observed line asymmetry across the acoustic power ridges is caused by correlated noise in observations \citep{nig98}, then it is possible that such correlated noise is an indirect cause of part of the observed phase shifts, although the magnitude and the detailed frequency dependence that this mechanism can incur remain to be investigated.

\section{Conclusion}
\label{sec5}

Through analyzing oscillatory signals observed at different optical depths, which roughly correspond to different atmospheric heights, we have found that the phases of evanescent acoustic waves continue to change with height, increasing for lower-frequency waves and decreasing for higher-frequency waves, and the changes are not accounted for by vertical flows. 
We also found that in areas with systematic upward or downward flows, the phase shifts measured in the acoustic waves are more substantial than those expectantly caused by flows. 
These pose great challenges to some helioseismic analyses that involve measurements using observations obtained at different atmospheric heights or in areas with systematic vertical flows. 
The comparison of the phases measured from IBIS and \sdo/HMI intensity data shows significant differences in magnitude, but these differences are likely owing to the different sensitivities of the instruments to high-$\ell$ waves.
We speculate that our measured frequency- and height-dependent phase changes in acoustic waves are probably due to the non-adiabaticity of the Sun's atmosphere with a combination of the asymmetry in the spectral line used in observations.
More investigations are needed to better quantify the phase changes as functions of height and frequency, so that this systematic effect can be either removed in helioseismic measuring processes or can be accounted for in helioseismic inversions. 

\begin{acknowledgments} 
We thank the anonymous referee for valuable comments that help improve the quality of this paper.
At the time the IBIS observation was acquired, Dunn Solar Telescope at Sacramento Peak, New Mexico was operated by National Solar Observatory (NSO).
NSO is operated by the Association of Universities for Research in Astronomy (AURA) Inc.~under a cooperative agreement with the National Science Foundation.
\sdo\ is a NASA mission, and HMI is an instrument developed by Stanford University under a NASA contract. 
We thank Dr.~M.~Waidele for carefully reading through the manuscript and providing useful comments.
J.Z.~was partly supported by a NASA grant 80NSSC19K0857. 
S.P.R.~acknowledges support from the Science and Engineering Research Board (SERB, Govt of India) grant CRG/2019/003786. 
\end{acknowledgments}

\end{document}